# Anomalous Co-site substitution effects on the physical properties of the thermoelectric oxide $NaCo_2O_4$

I. Terasaki

Department of Applied Physics, Waseda University, Tokyo 169-8555, JAPAN
and Precursory Research for Embryonic Science and Technology, Japan Science and Technology Co., JAPAN
e-mail: terra@mn.waseda.ac.jp    phone & fax: +81-3-5286-3854

**Abstract**

We prepared polycrystalline samples of $NaCo_{2-x}M_xO_4$ substituted for the Co site by other 3d or 4d elements (M=Mn, Cu, Zn, Ru, Rh, and Pd). While some of the elements (M=Mn, Ru and Rh) act as a strong scatterer to give rise to a rapid increase in resistivity, the others (M=Cu, Zn and Pd) do not deteriorate the electric conduction of the host. Most unusually, the Cu substitution significantly enhances the thermopower, which increases the figure of merit by several times at 100 K.

**Introduction**

Since many transition-metal oxides are harmless, abundant in earth, and stable in air at high temperature, they have been desired for an application to a thermoelectric generator of commercial use. This is, however, difficult to realize, because they are mostly poor conductors with a mobility of the order of 0.1-1 $cm^2/Vs$.

In 1997, we found a large thermopower (100 $\mu V/K$ at 300 K) accompanied by a low resistivity (200 $\mu\Omega cm$ at 300 K) for $NaCo_2O_4$ single crystals [1], and since then we have proposed thermoelectrics by oxide ceramics. We should note here that $NaCo_2O_4$ is an old material first synthesized in 1970's [2], and even its large thermopower was known in 1980's [3]. Thus our initial contribution might be only to show that a certain oxide can be a thermoelectric material. Perhaps the most important finding is that $NaCo_2O_4$ is more mysterious than we expected before: This material is also a poor conductor with a mobility of 0.1-1 $cm^2/Vs$ like other oxides, but instead the carrier concentration is two or three orders of magnitude larger than that of the state-of-the-art thermoelectric materials [4]. How such a large carrier concentration causes a large thermopower of 100 $\mu V/K$ is not trivial at all. We have proposed that the strong electron correlation plays an important role in the enhancement of the thermopower similarly to the case of heavy fermions [2-5]. In fact, the electron specific heat coefficient is found to be considerably large (40-50 $mJ/mol\ K^2$), which evidences a moderately strong correlation in this compound [5].

Another mystery of this compound is the Co-site substitution effects. As shown in Fig. 1, this compound is a layered oxide consisting of a $CoO_2$ block and a Na layer. The electric conduction is highly two-dimensional, and the resistivity anisotropy is 100-200 between the intra- and inter-layer directions [1]. Thus a small amount of substitution is expected to deteriorate the electric conduction. (Note that, according to the localization theory, a two-dimensional metal will be insulating at zero temperature in the presence of a finite amount of disorder). Contrary to this, Yakabe et al. [6] have found that the resistivity above room temperature is insensitive to some kinds of impurities such as Cu. Here we report on an extensive investigation of the substitution effects for $NaCo_2O_4$.

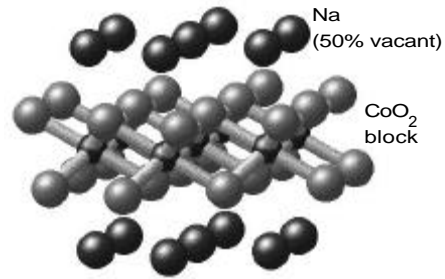

Fig.1 Crystal structure of $NaCo_2O_4$

**Experimental**

Polycrystalline samples of $Na_{1.2}Co_{2-x}M_xO_4$ (M=Mn, Cu, Zn, Ru, Rh, and Pd; x=0, 0.1, 0.2 and 0.3) were prepared through a solid-state reaction. A stoichiometric amount of $Na_2CO_3$, $Co_3O_4$, $Mn_2O_3$, CuO, ZnO, $RuO_2$, $Rh_2O_3$ and PdO was mixed and calcined at 1133 K for 12 h in air. The product was finely ground, pressed into a pellet, and sintered at 1193 K for 12 h in air. Since Na tends to evaporate during calcination, we added 20 % excess Na. Namely we expected samples of the nominal composition of $Na_{1.2}Co_{2-x}M_xO_4$ to be $NaCo_{2-x}M_xO_4$.

The resistivity was measured through a four-probe method from 4.2 to 300 K in a liquid He cryostat. The thermopower was measured using a steady-state technique from 4.2 to 300 K in a liquid He cryostat. Temperature gradient of 0.5 K/cm was generated by a small resistive heater pasted on one edge of the sample, and was monitored by a differential thermocouple made of copper-constantan. A thermopower of the voltage leads was carefully subtracted. The thermal conductivity was measured from 15 to 273 K using a steady-state technique in a closed refrigerator pumped down to $10^{-4}$ Pa. Temperature gradient was monitored by a differential thermocouple made of chromel-constantan.

The electronic states of the impurity were evaluated through a discrete variational X$\alpha$ (DVX$\alpha$) method, which was calculated though a program named SCAT [7]. A cluster of $Co_7O_{24}$ (the $CoO_2$ block in Fig.1) was used for the calculation of the $CoO_2$ block. To see the substitution effects, the Co atom in the center of the cluster was replaced by other 3d (4d) elements. Since the Na atoms are randomly occupies the regular sites, we tried various Na configurations, and found that the electronic states of Co 3d and O 2p do not change so much with the Na configuration. The density of states was plotted by broadening the energy levels of the cluster through a Gaussian function with a half width of 0.3 eV.

**Results and discussion**

Figure 2(a) shows the temperature dependence of the resistivity of $Na_{1.2}Co_{1.9}M_{0.1}O_4$ (M=Mn, Cu and Zn). The substituted Mn acts as a strong scatterer, which makes the resistivity one order of magnitude larger than the undoped one. This is indeed what is seen in other layered transition-metal oxides such as high-temperature superconducting cuprates, where only 3-5% Zn is enough to cause localization [8]. On the other hand, Zn and Cu do not appreciably affect resistivity. In particular, they do not increase the residual resistivity, which means that they have a negligibly small scattering cross section. This behavior is in a remarkable contrast with the Na-site substitution, where the resistivity monotonically increases with the substitution [9].

Figure 2(b) shows the temperature dependence of the thermopower of $Na_{1.2}Co_{1.9}M_{0.1}O_4$ (M=Mn, Cu and Zn). Normally a doped impurity increases only the scattering rate, and leaves the carrier concentration unchanged. Then, since the diffusive part of the thermopower (T-linear thermopower in metals) is a function of the Fermi energy, it is in principle independent of the amount of the doped impurities. Actually, the thermopower for Mn and Zn is nearly the same as that for the undoped sample. By contrast, the Cu substitution is quite

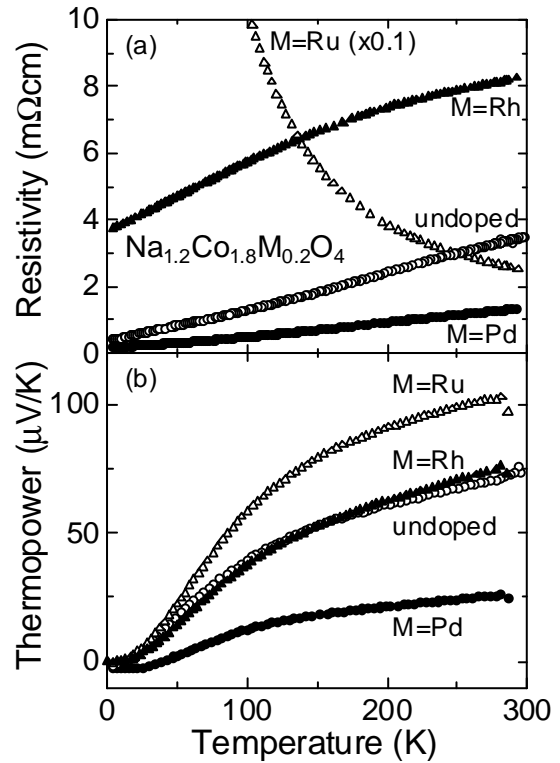

Fig. 3 (a) Resistivity and (b) thermopower of the $Na_{1.2}Co_{1.8}M_{0.2}O_4$ (M=Ru, Rh and Pd).

anomalous, and the magnitude is enhanced in all the measured temperature range from 4 to 300 K. This further shows that the Cu is surely substituted in the host: otherwise the thermopower would not be enhanced.

Let us proceed to the 4d substitution effects. Figure 3 shows the resistivity and thermopower of $Na_{1.2}Co_{1.8}M_{0.2}O_4$ (M=Ru, Rh and Pd). Since the wave function of 4d electrons is less localized than that of 3d electrons, the effects on the resistivity by the 4d impurity is milder than by the 3d impurities. M=Rh causes "usual" impurity effects, which increase the resistivity with the thermopower unchanged. For M=Ru, a decrease in the carrier concentration seems to occur in addition to the localization. Unusual impurity effects are seen in the Pd substitution: the sample for M=Pd has smaller resistivity and thermopower. The resistivity and thermopower decrease with increasing the Pd content up to 0.4. We observed the smallest thermopower of 5-6 μV/K at 300 K for $NaCo_{1.6}Pd_{0.4}O_4$. According to the scenario that the strong correlation is a main source of the large thermopower in $NaCo_2O_4$, Pd weakens the strong correlation to make the system a weakly correlated metal.

Figure 4 summarizes the thermoelectric properties of $Na_{1.2}Co_{2-x}Cu_xO_4$. As shown in Fig. 4(a), the resistivity (ρ) does not appreciably change with x. Reflecting that ρ is nearly independent of x, the thermal conductivity (κ) is also nearly independent of x as shown in Fig. 4(b). Note that the lattice thermal conductivity of $NaCo_2O_4$ is readily suppressed by the disordered Na layer. Since Na ions randomly occupy the 50% of the regular sites, the point-defect scattering by the

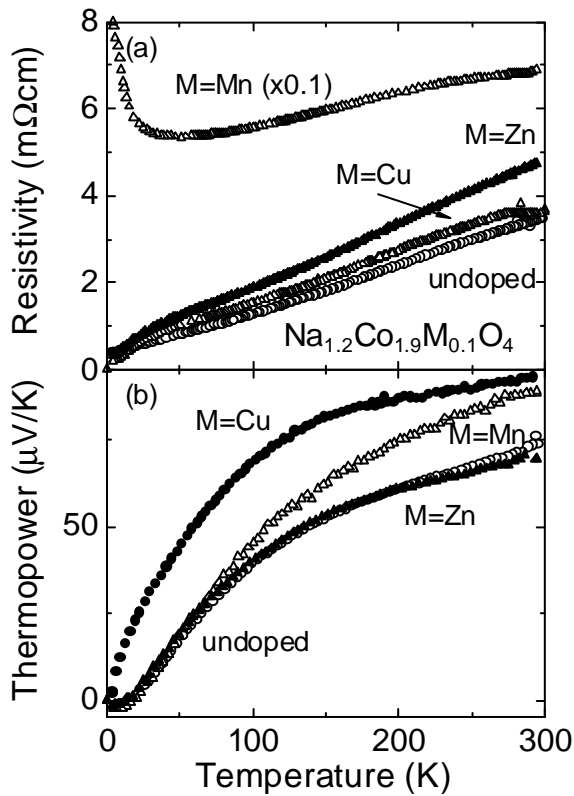

Fig. 2 (a) Resistivity and (b) thermopower of the $Na_{1.2}Co_{1.9}M_{0.1}O_4$ (M=Mn, Cu and Zn).

vacancies seriously reduces the lattice thermal conductivity [10]. Figure 4(c) exhibits a remarkable enhancement in the thermopower (S) for all the temperature measured in the present study. By combining ρ, κ, and S, the figure of merit $Z=S^2/\rho\kappa$ at 100 K for x=0.3 is found to be five times larger than that for x=0 as shown in Fig. 4(d). It should be pointed out that Cu is much more abundant and inexpensive than Co, which can be another merit for the Cu substitution from the viewpoint of commercial products.

At present how the substituted Cu and Pd modify the electronic states of $NaCo_{2-x}M_xO_4$ is an open question. We observed a red-shift of the reflectivity edge (plasma frequency) with x in the Cu substituted samples [11], and can say that the large thermopower is phenomenologically explained by the reduction of n/m, where n and m are the carrier concentration and the effective mass. A thermopower for a two-dimensional metal is given as S/T ~ m/n [12], and thus a small n/m will give a large thermopower.

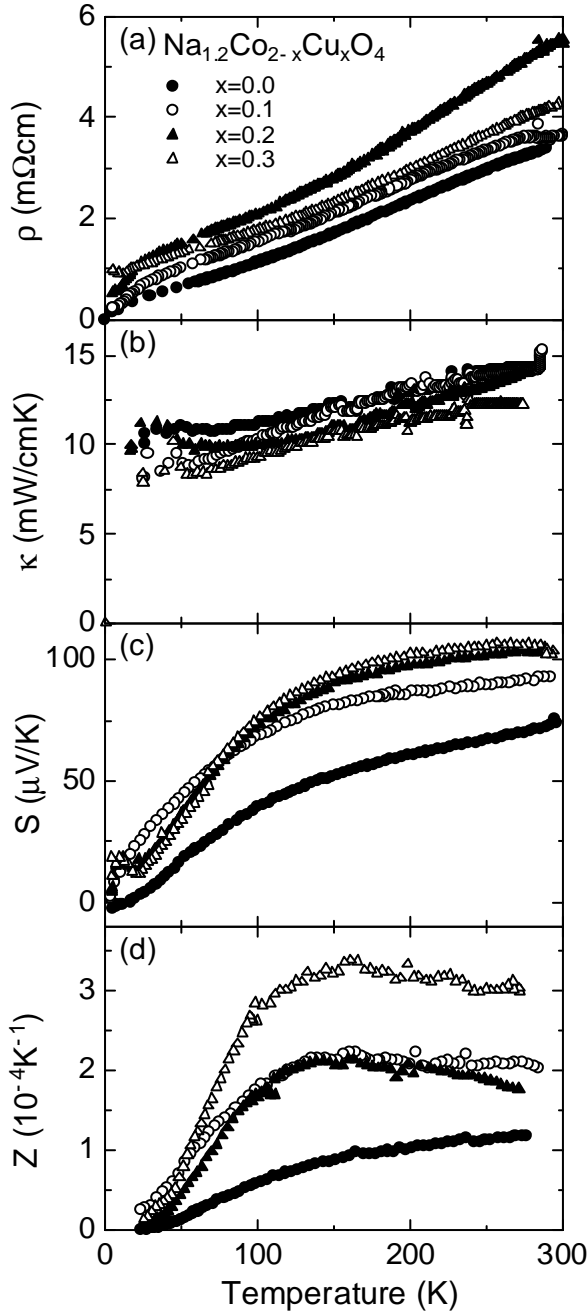

Fig. 4 Thermoelectric properties of $Na_{1.2}Co_{2-x}Cu_xO_4$.
(a) Resistivity, (b) thermal conductivity,
(c) thermopower and (d) figure of merit

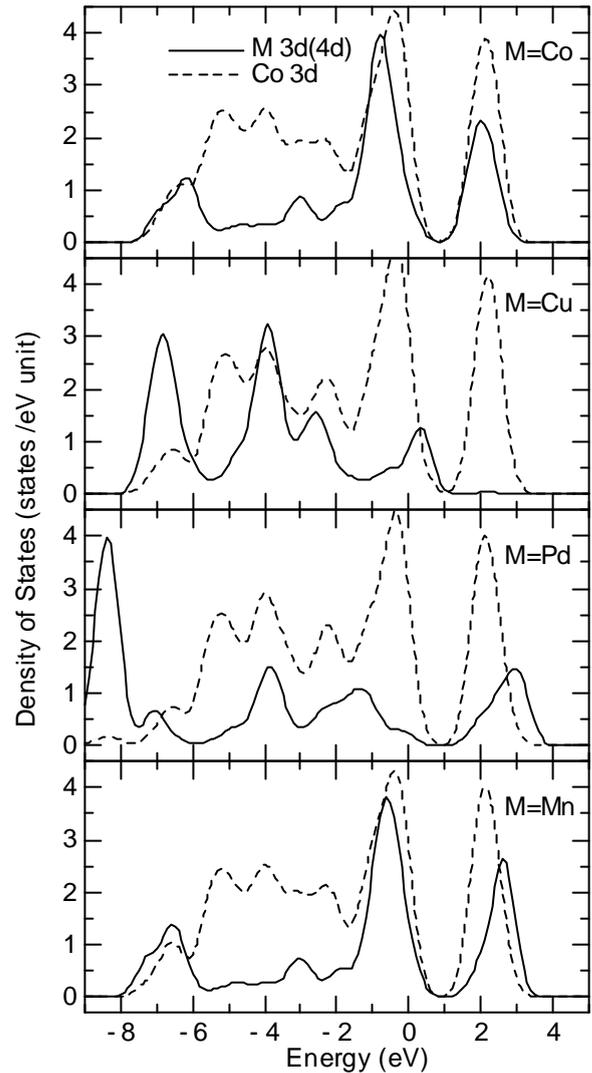

Fig. 5 The density of states calculated through DVXα method. The solid and dotted curves represent the impurity 3d (4d) levels, and the host Co 3d levels.

The DVXα calculation qualitatively reveals a difference between usual (Mn, Ru and Rh) and unusual (Cu, Zn and Pd) impurities. Figure 5 shows the calculated density of states of the 3d (4d) electrons for $Co_6MO_{24}$ clusters (M=Co, Cu, Pd and Mn). The M atom is in the center of the cluster, and is surrounded with the other outer six Co atoms. As shown by the dotted curves, the electronic states of the six Co atoms are nearly the same for all the calculations. Two large peaks are seen near –0.5 and 2 eV, which arises from the $t_{2g}$ and $e_g$ levels split due to the octahedrally coordinated oxygen. Deeper peaks result from the hybridization with oxygen 2p states. For M=Co, the dotted and solid curves are essentially similar from -2 to 3 eV. This indicates that the electronic states of Co 3d is essentially similar between the center and outer atoms, and that the self consistency of the calculation is satisfactory. It should be emphasized that the calculated density of states are quantitatively consistent with the local-density-approximation band calculation by Singh [13].

For M=Cu and Pd, the density of states near 0 eV is quite small, and most of the energy levels of Cu 3d and Pd 4d are found to be much deeper than Co 3d of the host. This implies that the impurity levels are much deeper than the thermal energy $k_BT$, which might give a small scattering cross section. On the contrary, the energy levels of Mn 3d are very close to those of Co 3d of the host, which would affect the electric conduction. These difference might come from the number of d electrons. For example, a Cu atom tends to exist in oxides as $Cu^{2+}$ or $Cu^+$, which has the configuration of $(3d)^9$ or $(3d)^{10}$. In either case, the $t_{2g}$ levels are filled and located at deeper energies. Then they will not hybridize the highest molecular orbital of the host Co 3d of $t_{2g}$ symmetry.

Singh [13] has found, from his calculated band structure of $NaCo_2O_4$, that the valence band of this compound is extremely narrow. In other words, the band mass of $NaCo_2O_4$ is extremely large (and the mass enhancement due to the electron correlation is rather moderate), which predicts the room-temperature thermopower of 110 μV/K. We wonder if his scenario could explain the Cu- and Pd-substitution effects. Usually an electron with a larger band mass tends to be localized by a smaller amount of disorder, and is unlikely to be metallic against the 15 % substitution of the impurities.

If the Cu enhanced the electron correlation (though we cannot understand the mechanism), m and τ would be enhanced in the same way to keep mobility τ/m unchanged, where τ is the scattering time. In this case S would be enhanced, while ρ and κ remain intact. If so, the electron correlation of $NaCo_2O_4$ can be controlled by the Co-site substitution of Cu.

**Summary**

A set of polycrystalline samples of the thermoelectric oxide $Na_{1.2}Co_{2-x}M_xO_4$ (M=Mn, Cu, Zn, Ru, Rh, and Pd; x=0-0.3) were prepared. The Cu- and Pd-substitutions are quite anomalous, and are seriously incompatible with conventional impurity effects. With increasing Cu content, the thermopower remarkably enhances particularly at low temperatures, while the resistivity and thermal conductivity essentially remain unchanged. As a result, the thermoelectric figure of merit for x=0.3 at 100 K is five times larger than that for x=0. On the other hand, both the resistivity and thermopower decreases with Pd concentration. If a main source of the large thermopower is the strong electron correlation, the present results imply that some kinds of substitution can control the strength of the correlation in this compound.

**Acknowledgments**

I would like to thank Y. Ando, K. Kohn, S. Kurihara, T. Koyanagi, W. Shin, N. Murayama, and K. Koumoto for fruitful discussion. I would also like to thank my graduate and undergraduate students, Y. Ishii, R. Kitawaki, D. Tanaka, K. Takahata and Y. Iguchi for collaboration.